\documentclass[pra,aps,nofootinbib,notitlepage,superscriptaddress,twocolumn]{revtex4-1}
\usepackage{amsthm}
\usepackage{amsmath,bm}
\usepackage{amssymb}
\usepackage{amsfonts}
\usepackage{graphicx}
\usepackage{fancyhdr}
\usepackage{txfonts}
\usepackage[vcentermath]{youngtab}
\usepackage[colorlinks=true,linkcolor=blue,citecolor=blue,urlcolor=blue]{hyperref}

\begin{document}
\preprint{HEP/123-qed}
\title{Generalized Bell-like inequality and maximum violation for multiparticle
entangled Schr\"{o}dinger-cat-states of spin-s}
\author{Yan Gu}
\author{Wei-Dong Li}
\author{Xiao-Lei Hao}
\author{Jiu-Qing Liang}
\email{jqliang@sxu.edu.cn}
\affiliation{Institute of Theoretical Physics and Department of Physics, State Key
Laboratory of Quantum Optics and Quantum Optics Devices, Shanxi University,
Taiyuan, Shanxi 030006, China.}
\author{Lian-Fu Wei}
\affiliation{Information Quantum Technology Laboratory, International Cooperation Research
Center of China Communication and Sensor Networks for Modern Transportation,
School of Information Science and Technology, Southwest Jiaotong University,
Chengdu 610031, China}

\begin{abstract}
This paper proposes a generalized Bell-like inequality (GBI) for multiparticle
entangled Schr\"{o}dinger-cat--states of arbitrary spin-$s$. Based on quantum
probability statistics the GBI and violation are formulated in an unified
manner with the help of state density operator, which can be separated to
local and non-local parts. The local part gives rise to the inequality, while
the non-local part is responsible for the violation. The GBI is not violated
at all by quantum average except the spin-$1/2$ entangled states. If the
measuring outcomes are restricted in the subspace of spin coherent state
(SCS), namely, only the maximum spin values $\pm s$, the GBI is still
meaningful for the incomplete measurement. With the help of SCS quantum
probability statistics, it is proved that the violation of GBI can occur only
for half-integer spins but not integer spins. Moreover, the maximum violation
bound depends on the number parity of entangled particles, that it is $1/2$
for the odd particle-numbers while $1$ for even numbers.

\end{abstract}
\volumeyear{year}
\volumenumber{number}
\issuenumber{number}
\eid{identifier}
\date[Date text]{date}
\received[Received text]{date}

\revised[Revised text]{date}

\accepted[Accepted text]{date}

\published[Published text]{date}

\startpage{1}
\endpage{2}
\maketitle
\preprint{ }



\thispagestyle{fancy}

\section{Introduction}

Non-locality\cite{non1,non2,non3} is regarded as the most peculiar
characteristic of quantum mechanics, since it does not coexist with the
relativistic causality in our intuition of space and time. Quantum
entangled-state originally was introduced by Einstein-Podolsky-Rosen as a
critical example for the non-locality\cite{EPR} showing apparently
contradictory results with the locality and reality criterion in classical
theory. The quantum entanglement now has become a key concept of quantum
information and computation\cite{inf1,inf2,inf3,inf4,inf5}.\textbf{\ }Quantum
correlations between entangled systems are fundamentally different from
classical correlations\cite{class1,class2}, especially when these systems are
spatially separated. It was Bell who proposed for the first time an inequality
known as Bell inequality (BI) to test this difference\cite{Bell}. The BI is
actually a constraint on the correlations compatible with local
hidden-variable (or local realistic) theories.\textbf{\ }Great attentions have
been paid on theoretical and experimental studies of the
inequality\cite{e1,e2,e3,e4,e5}. Many\textbf{\ }experimental
evidences\cite{v3,v4,v5,v6,v7,v8,v9} confirm the violation of BI, which
provides an overwhelming superiority for the non-locality in quantum
mechanics\cite{m1,m2}.

Stimulated by original work of Bell various extensions were proposed such as
Clauser-Horne-Shimony-Holt (CHSH)\cite{CHSH} inequality and Wigner inequality
(WI)\cite{WI1}, in which\textbf{\ }the particle number probability of positive
spin is measured only\cite{WI1,WI2}.\textbf{\ }A loophole-free experiment was
reported recently to verify the violation of CHSH inequality with the
electronic spin of nitrogen-vacancy defect in a diamond
chip\cite{diamond1,diamond2}. The violation was also confirmed experimentally
with\textbf{\ }two-photon entangled states of mutually perpendicular
polarizations\cite{photon}. By closing the two main loopholes ( the
\textquotedblleft locality loophole\textquotedblright\ and \textquotedblleft
detection loophole\textquotedblright\ ) at the same time, some teams have
independently confirmed that we must definitely renounce local
realism\cite{diamond1,close1,close2,close3}.\textbf{\ }

Theoretical analysis for the violation of BI was presented in the beginning of
90's\cite{Gisin,pope}. In terms of spin-coherent-state (SCS) quantum
probability statistics the BI, WI and their maximum violation-bounds are
formulated uniformly for arbitrary two-spin entangled states with antiparallel
and parallel spin polarizations\cite{p1,p2,p3,p4,p5}. The density operator of
entangled state is separated into the \textquotedblleft
local\textquotedblright\ (classical probability state) and \textquotedblleft
non-local\textquotedblright\ (quantum interference between two components of
entangled state) parts. The local part gives rise to the BI or WI, while the
non-local part is responsible for the violation.

The Bell correlation for two-particle entangled state of arbitrary spin-$s$
called the Schr\"{o}dinger cat-state is also investigated with the SCS quantum
probability statistics. The BI is not violated by the entangled
Schr\"{o}dinger cat-states at all except the spin-$1/2$ case\cite{p5}. If, on
other hand, the measuring outcomes are confined in the subspace of SCS, a
universal BI is formulated in terms of the \textquotedblleft
local\textquotedblright\ part of density operator.\textbf{\ }The maximum
violation bound is found for the entangled cat states with both antiparallel
and parallel polarizations\cite{p1,p2}. Particularly\textbf{\ }a spin parity
effect\cite{p5} is observed that the universal BI can be violated only by the
entangled cat states of half-integer but not the integer spins.\textbf{\ }The
violation of universal BI is seen to be a direct result of non-trivial Berry
phase between the SCSs\cite{p1,p2,p3,p4,p5} of south- and north-pole gauges
for half-integer spin,\textbf{\ }while the geometric phase is trivial for the
integer spins. This observation\cite{p5} provides an example to relate the
violation of BI with the geometric phase.

We in the present paper study the multiparticle entangled cat-states of
arbitrary spin-$s$, which just are the well known Greenberger, Horne, and
Zeilinger (GHZ)\cite{GHZ1,GHZ2,GHZ3} state in the spin-$1/2$
case\cite{cat,cat2}.{ }The quantum entanglement was originally generalized to
GHZ state for four spin-$1/2$ particles, while later was simplified to three
spin-$1/2$ particles\cite{s3}, and was verified experimentally\cite{s4,s5}.{
}The violations of inequalities for $n$ spin-$1/2$ particles have been studied
extensively\cite{ex1,ex2,ex3,ex4,ex5,ex6,ex7,ex8,ex9,ex10,ach7}. Bell's
inequality has been proposed for $n$ spin-$s$ particles by $n$ distant
observers\cite{mk}.

The main goal of the present study is to extend the two-particle universal BI
to multiparticle entangled cat-state of arbitrary spin-$s$. The entanglement
in many-particle systems has been thoroughly investigated with significant
progress achieved\cite{ach1,ach2,ach3,ach4,ach5,ach6,ach7,ach8}. The
many-particle non-local correlation plays an important role in phase
transitions and criticality in condensed matters\cite{phase}.\textbf{\ }It
might also enhance our understanding of entanglement application in quantum
information theory.

Although the non-local correlation for two-particle entangled cat-state is
investigated, a suitable inequality for arbitrary many-particle state has not
be found in our previous paper\cite{p5}. In the present paper, a generalized
Bell-like inequality (GBI) is discovered to characterize the non-locality of
multiparticle entangled cat-state with arbitrary spin-$s$. The GBI and the
maximum violation bound are formulated in an unified formalism with SCS
quantum probability statistics.

In Section II the GBI and its maximum violation bound are formulated for
$n$-particle entangled states of spin-$1/2$. It is demonstrated in Sec.III
that the GBI and the maximum violation exist for the $n$-particle entangled
cat-states if the measuring outcomes are restricted in the subspace of SCS.
Moreover we observe interesting spin and particle number parity effects in the
violation of GBI.

\section{GBI and maximum violation for $n$-particle entangled state of
spin-$1/2$}

For the two-particle entangled state of spin-$s$ it is proved that the
non-local correlation vanishes\cite{p5} $P_{s}^{nlc}(a,b)=0$ except the case
of $s=1/2$, since the average transition vanishes $\langle\pm s|\hat{S}%
\cdot\mathbf{n|\mp}s\rangle=0$ ($\mathbf{n=a,b}$) induced by the spin
projection-operator. The BI is not violated at all. We then consider the
measurements confined in the subspace of SCS, namely only the maximum spin
values $\pm s$ are measured\cite{p5}. In this case the original BI is no
longer valid, because the total measuring-outcome probability is less than
one. The universal BI is suitable for complete and partial measurement. It is
valid also for entangled states with both antiparallel and parallel spin
polarizations. A spin parity effect is observed that the universal BI can be
violated only by the entangled cat-states of half-integer but not the integer
spins. A maximum violation bound of universal BI is found as $p_{s}^{\max}=1$,
which is valid for arbitrary half-integer spin-$s$ states. In the present
paper, we consider the $n$-particle entangled cat-state of spin-$s$%
\begin{equation}
\left\vert \psi\right\rangle =c_{1}|+s\rangle^{\otimes n}+c_{2}|-s\rangle
^{\otimes n}, \label{1}%
\end{equation}
in which $c_{1}=e^{i\eta}\sin\xi,c_{2}=e^{-i\eta}\cos\xi$ characterize the
normalized coefficient with two arbitrary real-parameters $\xi,\eta$. The
state density operator can be separated into the local and non-local part%
\[
\hat{\rho}=\hat{\rho}_{lc}+\hat{\rho}_{nlc}%
\]
with
\[
\hat{\rho}_{lc}=\sin^{2}\xi|+s\rangle^{\otimes n}\left\langle +s\right\vert
^{\otimes n}+\cos^{2}\xi|-s\rangle^{\otimes n}\left\langle -s\right\vert
^{\otimes n},
\]%
\[
\hat{\rho}_{nlc}=\sin\xi\cos\xi\left(  e^{i2\eta}|+s\rangle^{\otimes
n}\left\langle -s\right\vert ^{\otimes n}+e^{-i2\eta}|-s\rangle^{\otimes
n}\left\langle +s\right\vert ^{\otimes n}\right)  .
\]

The normalized outcome correlation from $n$ observers can be also separated as
the local and non-local parts
\begin{align}
p(a_{1},a_{2},...,a_{n})  &  =\frac{1}{s^{n}}Tr\left[  \hat{\rho}\hat{\Omega
}(a_{1},a_{2},...,a_{n})\right] \nonumber\\
&  =p_{lc}\left(  a_{1},a_{2},...,a_{n}\right)  +p_{nlc}\left(  a_{1}%
,a_{2},...,a_{n}\right)  , \label{p}%
\end{align}
measured respectively along $a_{1},a_{2},...,a_{n}$ directions. The measuring
correlation operator is
\[
\hat{\Omega}(a_{1},a_{2},...,a_{n})=\left(  \hat{s}\cdot\mathbf{a}_{1}\right)
\otimes\left(  \hat{s}\cdot\mathbf{a}_{2}\right)  \otimes...\otimes\left(
\hat{s}\cdot\mathbf{a}_{n}\right)  .
\]

The universal BI for the two-particle entangled state\cite{p5} is then
extended directly to a GBI for $n$-particle entangled cat-state of spin-$s$
Eq.(\ref{1})
\begin{align}
&  p_{lc}\left(  a_{1},a_{2},...,a_{n}\right)  p_{lc}\left(  a_{2}%
,a_{3},...,a_{n+1}\right)  p_{lc}\left(  a_{3},a_{4},...,a_{n+2}\right)
\nonumber\\
&  \times...\times p_{lc}\left(  a_{n},a_{n+1},...,a_{m}\right) \nonumber\\
&  \leq\left\vert p_{lc}\left(  a_{1},a_{3},...,a_{m}\right)  \right\vert ,
\label{2}%
\end{align}
with total measuring directions $m=2n-1$. The validity of GBI Eq.(\ref{2}) is
obviously by means of hidden variable classical-statistics\cite{p5} following
Bell\cite{Bell}, since any two-particle normalized measuring-outcomes (denoted
by $A=\pm1$ and $B=\pm1$) have the relation
$\vert$%
$A$($a$)%
$\vert$%
$=$
$\vert$%
$B$($a$)%
$\vert$
along the same direction $a$ and $A^{2}(a)=$ $B^{2}(a)=1$. The derivation of
GBI Eq.(\ref{2}) is presented in Appendix.

\subsection{Three-particle entangled state of spin-1/2}

As a simple example we consider the three-particle entangled state of
spin-$1/2$
\[
\left\vert \psi\right\rangle =e^{i\eta}\sin\xi|+,+,+\rangle+e^{-i\eta}\cos
\xi|-,-,-\rangle,
\]
where $\left\vert +\right\rangle $ and $\left\vert -\right\rangle $ are the
spin-up and -down states along the $z$ axis. The local part of density
operator $\hat{\rho}=\left\vert \psi\right\rangle \langle\psi|$%
\[
\hat{\rho}_{lc}=\sin^{2}\xi|+,+,+\rangle\left\langle +,+,+\right\vert
+\cos^{2}\xi|-,-,-\rangle\left\langle -,-,-\right\vert
\]
results in the GBI. While the non-local part%
\[
\hat{\rho}_{nlc}=\sin\xi\cos\xi\left(
\begin{array}
[c]{c}%
e^{i2\eta}|+,+,+\rangle\left\langle -,-,-\right\vert \\
+e^{-i2\eta}|-,-,-\rangle\left\langle +,+,+\right\vert
\end{array}
\right)  ,
\]
which describes the quantum interference\cite{p1,p2,p3,p4,p5} between
two-component of the entangled state, leads to the violation of the GBI.

\subsubsection{Verification of GBI with the local part of correlation}

Let us suppose that three spins are measured independently by three observers
along arbitrary directions $\mathbf{a}_{1}$, $\mathbf{a}_{2}$ and
$\mathbf{a}_{3}$. The measuring outcomes of each spin fall into the
eigenvalues of projection spin-operator $\hat{s}\cdot\mathbf{r}$ $\left(
\mathbf{r}=\mathbf{a}_{1},\mathbf{a}_{2},\mathbf{a}_{3}\right)  $, i.e.
\begin{equation}
\hat{s}\cdot\mathbf{r}\left\vert \pm\mathbf{r}\right\rangle =\mathbf{\pm}%
\frac{1}{2}\left\vert \pm\mathbf{r}\right\rangle , \label{3}%
\end{equation}
(in the unit convention $\hbar=1$) for the case $s=1/2$. The unit vector of
arbitrary direction $\mathbf{r}=\left(  \sin\theta_{r}\cos\phi_{r},\sin
\theta_{r}\sin\phi_{r},\cos\theta_{r}\right)  $ is parameterized with the
polar and azimuthal angles $\theta_{r},$ $\phi_{r}$ in spherical coordinates.
Two eigenstates of the Eq.(\ref{3}) are given by%
\begin{align}
\left\vert +\mathbf{r}\right\rangle  &  =\cos\frac{\theta_{r}}{2}\left\vert
+\right\rangle +\sin\frac{\theta_{r}}{2}e^{i\phi_{r}}\left\vert -\right\rangle
,\nonumber\\
\left\vert -\mathbf{r}\right\rangle  &  =\sin\frac{\theta_{r}}{2}\left\vert
+\right\rangle -\cos\frac{\theta_{r}}{2}e^{i\phi_{r}}\left\vert -\right\rangle
, \label{eig}%
\end{align}
which are known as the spin coherent states of north- and south-pole
gauges\cite{p1,p2,p3,p4}. The independent measuring-outcome basis vectors are
labeled as%
\begin{align}
\left\vert 1\right\rangle  &  =|+\mathbf{a}_{1},+\mathbf{a}_{2},+\mathbf{a}%
_{3}\rangle,\left\vert 2\right\rangle =|+\mathbf{a}_{1},-\mathbf{a}%
_{2},-\mathbf{a}_{3}\rangle,\nonumber\\
\left\vert 3\right\rangle  &  =|-\mathbf{a}_{1},+\mathbf{a}_{2},-\mathbf{a}%
_{3}\rangle,\left\vert 4\right\rangle =|-\mathbf{a}_{1},-\mathbf{a}%
_{2},+\mathbf{a}_{3}\rangle,\nonumber\\
\left\vert 5\right\rangle  &  =|+\mathbf{a}_{1},+\mathbf{a}_{2},-\mathbf{a}%
_{3}\rangle,\left\vert 6\right\rangle =|+\mathbf{a}_{1},-\mathbf{a}%
_{2},+\mathbf{a}_{3}\rangle,\nonumber\\
\left\vert 7\right\rangle  &  =|-\mathbf{a}_{1},+\mathbf{a}_{2},+\mathbf{a}%
_{3}\rangle,\left\vert 8\right\rangle =|-\mathbf{a}_{1},-\mathbf{a}%
_{2},-\mathbf{a}_{3}\rangle, \label{base}%
\end{align}
for the sake of simplicity. The basis vectors are the eigenstates of the
correlation operator such that
\[
\hat{\Omega}\left(  a_{1},a_{2},a_{3}\right)  \left\vert i\right\rangle
=\pm\left(  \frac{1}{2}\right)  ^{3}\left\vert i\right\rangle ,
\]
respectively for $i=1,2,3,4$ and $5,6,7,8$. Thus we have the (normalized)
total correlation probability evaluated from the matrix elements of density
operator only
\[
p(a_{1},a_{2},a_{3})=\sum_{i=1}^{4}\rho_{ii}-\sum_{i=5}^{8}\rho_{ii}.
\]
From the local matrix elements of density operator
\[
\rho_{11}^{lc}=\sin^{2}\xi%
{\displaystyle\prod\limits_{i=1}^{3}}
K_{a_{i}}^{2}+\cos^{2}\xi%
{\displaystyle\prod\limits_{i=1}^{3}}
\Gamma_{a_{i}}^{2},
\]%
\[
\rho_{22}^{lc}=\sin^{2}\xi K_{a_{1}}^{2}\Gamma_{a_{2}}^{2}\Gamma_{a_{3}}%
^{2}+\cos^{2}\xi\Gamma_{a_{1}}^{2}K_{a_{2}}^{2}K_{a_{3}}^{2},
\]%
\[
\rho_{33}^{lc}=\sin^{2}\xi\Gamma_{a_{1}}^{2}K_{a_{2}}^{2}\Gamma_{a_{3}}%
^{2}+\cos^{2}\xi K_{a_{1}}^{2}\Gamma_{a_{2}}^{2}K_{a_{3}}^{2},
\]%
\[
\rho_{44}^{lc}=\sin^{2}\xi\Gamma_{a_{1}}^{2}\Gamma_{a_{2}}^{2}K_{a_{3}}%
^{2}+\cos^{2}\xi K_{a_{1}}^{2}K_{a_{2}}^{2}\Gamma_{a_{3}}^{2},
\]%
\[
\rho_{55}^{lc}=\sin^{2}\xi K_{a_{1}}^{2}K_{a_{2}}^{2}\Gamma_{a_{3}}^{2}%
+\cos^{2}\xi\Gamma_{a_{1}}^{2}\Gamma_{a_{2}}^{2}K_{a_{3}}^{2},
\]%
\[
\rho_{66}^{lc}=\sin^{2}\xi K_{a_{1}}^{2}\Gamma_{a_{2}}^{2}K_{a_{3}}^{2}%
+\cos^{2}\xi\Gamma_{a_{1}}^{2}K_{a_{2}}^{2}\Gamma_{a_{3}}^{2},
\]%
\[
\rho_{77}^{lc}=\sin^{2}\xi\Gamma_{a_{1}}^{2}K_{a_{2}}^{2}K_{a_{3}}^{2}%
+\cos^{2}\xi K_{a_{1}}^{2}\Gamma_{a_{2}}^{2}\Gamma_{a_{3}}^{2},
\]%
\begin{equation}
\rho_{88}^{lc}=\sin^{2}\xi%
{\displaystyle\prod\limits_{i=1}^{3}}
\Gamma_{a_{i}}^{2}+\cos^{2}\xi%
{\displaystyle\prod\limits_{i=1}^{3}}
K_{a_{i}}^{2}, \label{element}%
\end{equation}
in which $K_{r}=\cos\theta_{r}/2,\Gamma_{r}=\sin\theta_{r}/2$ for
$r=a_{1},a_{2},a_{3}$, the local part of correlation is found as%
\begin{equation}
p_{lc}(a_{1},a_{2},a_{3})=-\cos\left(  2\xi\right)
{\displaystyle\prod\limits_{i=1}^{3}}
\cos\theta_{a_{i}}. \label{plc}%
\end{equation}
The GBI of Eq.(\ref{2}) for $n=3$ can be verified with the local part of
correlation probability Eq.(\ref{plc}) such that
\begin{align*}
&  p_{lc}\left(  a_{1},a_{2},a_{3}\right)  p_{lc}\left(  a_{2},a_{3}%
,a_{4}\right)  p_{lc}\left(  a_{3},a_{4},a_{5}\right) \\
&  =\cos^{2}\left(  2\xi\right)  \left(
{\displaystyle\prod\limits_{i=2}^{4}}
\cos^{2}\theta_{a_{i}}\right)  p_{lc}\left(  a_{1},a_{3},a_{5}\right) \\
&  \leq\left\vert p_{lc}\left(  a_{1},a_{3},a_{5}\right)  \right\vert .
\end{align*}

\subsubsection{Maximum violation of GBI}

The non-local parts of density operator are evaluated as%
\[
\rho_{11}^{nlc}=\frac{1}{2^{3}}\sin\left(  2\xi\right)  \left(
{\displaystyle\prod\limits_{i=1}^{3}}
\sin\theta_{a_{i}}\right)  \cos\left(
{\displaystyle\sum\limits_{i=1}^{3}}
\phi_{a_{i}}+2\eta\right)  ,
\]
with%
\[
\rho_{ii}^{nlc}=\rho_{11}^{nlc},
\]
for $i=2,3,4$ and%
\[
\rho_{jj}^{nlc}=-\rho_{11}^{nlc}.
\]
for $j=5,6,7,8$. The non-local part of correlation is%
\begin{equation}
p_{nlc}(a_{1},a_{2},a_{3})=\sin\left(  2\xi\right)  \left(
{\displaystyle\prod\limits_{i=1}^{3}}
\sin\theta_{a_{i}}\right)  \cos\left(
{\displaystyle\sum\limits_{i=1}^{3}}
\phi_{a_{i}}+2\eta\right)  . \label{pnlc}%
\end{equation}
The entire (normalized) quantum correlation-probability becomes
\begin{align}
&  p(a_{1},a_{2},a_{3})\nonumber\\
&  =-\cos\left(  2\xi\right)  \left(
{\displaystyle\prod\limits_{i=1}^{3}}
\cos\theta_{a_{i}}\right) \nonumber\\
&  +\sin\left(  2\xi\right)  \left(
{\displaystyle\prod\limits_{i=1}^{3}}
\sin\theta_{a_{i}}\right)  \cos\left(
{\displaystyle\sum\limits_{i=1}^{3}}
\phi_{a_{i}}+2\eta\right)  . \label{qcorr}%
\end{align}
In order to find the maximum violation of the GBI we define a quantum
correlation-probability difference
\begin{align}
p_{GB}  &  =p\left(  a_{1},a_{2},a_{3}\right)  p\left(  a_{2},a_{3}%
,a_{4}\right)  p\left(  a_{3},a_{4},a_{5}\right) \nonumber\\
&  -\left\vert p\left(  a_{1},a_{3},a_{5}\right)  \right\vert . \label{diff}%
\end{align}
The GBI becomes%
\begin{equation}
p_{GB}^{lc}\leq0. \label{GBI}%
\end{equation}
Thus any positive value of $p_{GB}$ indicates the violation of GBI. The
maximum violation appears with the polar angles $\theta_{a_{1}}=\theta_{a_{2}%
}=\theta_{a_{3}}=\pi/2,$ where the local part of correlation vanishes. Thus
the correlation-probability is simplified as%
\[
p(a_{1},a_{2},a_{3})=\sin\left(  2\xi\right)  \cos\left(
{\displaystyle\sum\limits_{i=1}^{3}}
\phi_{a_{i}}+2\eta\right)  .
\]
The probability difference Eq.(\ref{diff}) reads%
\begin{align*}
p_{GB}  &  =\sin^{3}\left(  2\xi\right)  \cos\left(
{\displaystyle\sum\limits_{i=1}^{3}}
\phi_{a_{i}}+2\eta\right) \\
&  \times\cos\left(
{\displaystyle\sum\limits_{i=2}^{4}}
\phi_{a_{i}}+2\eta\right)  \cos\left(
{\displaystyle\sum\limits_{i=3}^{5}}
\phi_{a_{i}}+2\eta\right) \\
&  -\left\vert \sin\left(  2\xi\right)  \cos\left(
{\displaystyle\sum\limits_{i=0}^{2}}
\phi_{a_{2i+1}}+2\eta\right)  \right\vert ,
\end{align*}
which becomes
\begin{align*}
p_{GB}  &  =-\sin\left(
{\displaystyle\sum\limits_{i=1}^{3}}
\phi_{a_{i}}\right)  \sin\left(
{\displaystyle\sum\limits_{i=2}^{4}}
\phi_{a_{i}}\right)  \sin\left(
{\displaystyle\sum\limits_{i=3}^{5}}
\phi_{a_{i}}\right) \\
&  -\left\vert \sin\left(
{\displaystyle\sum\limits_{i=0}^{2}}
\phi_{a_{2i+1}}\right)  \right\vert ,
\end{align*}
for the state parameters $\xi=\eta=\pi/4\mathrm{mod}2\pi$. With the azimuthal
angles of $5$ measuring directions\textbf{ }$\phi_{a_{2i+1}}=0$ $(i=0,1,2)$,
$\phi_{a_{2}}=\phi_{a_{4}}=3\pi/4,$ we have the maximum violation
\begin{equation}
p_{GB}^{\max}=\frac{1}{2}. \label{max3}%
\end{equation}

\subsection{Four-particle entangled state of spin-1/2}

For the four-particle entangled state there are $16$ independent basis vectors
for the arbitrary measuring directions denoted by $\mathbf{a}_{1}%
,\mathbf{a}_{2},\mathbf{a}_{3}$ and $\mathbf{a}_{4}$:%
\begin{align}
\left\vert 1\right\rangle  &  =|\mathbf{+a}_{1},+\mathbf{a}_{2},+\mathbf{a}%
_{3}\mathbf{,+a}_{4}\rangle,\left\vert 2\right\rangle =\mathbf{|\mathbf{+a}%
}_{1}\mathbf{,+\mathbf{a}}_{2}\mathbf{,-\mathbf{a}}_{3}\mathbf{\mathbf{,-a}%
}_{4}\mathbf{\rangle},\nonumber\\
\left\vert 3\right\rangle  &  =\mathbf{|\mathbf{+a}}_{1}\mathbf{,-\mathbf{a}%
}_{2}\mathbf{,+\mathbf{a}}_{3}\mathbf{\mathbf{,-a}}_{4}\mathbf{\rangle
},\left\vert 4\right\rangle =\mathbf{|\mathbf{+a}}_{1}\mathbf{,-\mathbf{a}%
}_{2}\mathbf{,-\mathbf{a}}_{3}\mathbf{\mathbf{,+a}}_{4}\mathbf{\rangle
},\nonumber\\
\left\vert 5\right\rangle  &  =\mathbf{|-\mathbf{a}}_{1}\mathbf{,+\mathbf{a}%
}_{2}\mathbf{,+\mathbf{a}}_{3}\mathbf{\mathbf{,-a}}_{4}\mathbf{\rangle
},\left\vert 6\right\rangle =\mathbf{|-\mathbf{a}}_{1}\mathbf{,+\mathbf{a}%
}_{2}\mathbf{,-\mathbf{a}}_{3}\mathbf{\mathbf{,+a}}_{4}\mathbf{\rangle
},\nonumber\\
\left\vert 7\right\rangle  &  =|\mathbf{\mathbf{-a}}_{1}\mathbf{,-\mathbf{a}%
}_{2}\mathbf{,+\mathbf{a}}_{3}\mathbf{\mathbf{,+a}}_{4}\rangle,\left\vert
8\right\rangle =|\mathbf{\mathbf{-a}}_{1}\mathbf{,-\mathbf{a}}_{2}%
\mathbf{,-\mathbf{a}}_{3}\mathbf{\mathbf{,-a}}_{4}\rangle,\nonumber\\
\left\vert 9\right\rangle  &  =\mathbf{|\mathbf{+a}}_{1}\mathbf{,+\mathbf{a}%
}_{2}\mathbf{,+\mathbf{a}}_{3}\mathbf{\mathbf{,-a}}_{4}\mathbf{\rangle
},\left\vert 10\right\rangle =\mathbf{|\mathbf{+a}}_{1}\mathbf{,+\mathbf{a}%
}_{2}\mathbf{,-\mathbf{a}}_{3}\mathbf{\mathbf{,+a}}_{4}\mathbf{\rangle
},\nonumber\\
\left\vert 11\right\rangle  &  =\mathbf{\mathbf{|\mathbf{+a}}}_{1}%
\mathbf{\mathbf{,-\mathbf{a}}}_{2}\mathbf{\mathbf{,+\mathbf{a}}}%
_{3}\mathbf{\mathbf{\mathbf{,+a}}}_{4}\mathbf{\mathbf{\rangle},}\left\vert
12\right\rangle =\mathbf{|\mathbf{+a}}_{1}\mathbf{,-\mathbf{a}}_{2}%
\mathbf{,-\mathbf{a}}_{3}\mathbf{\mathbf{,-a}}_{4}\mathbf{\rangle},\nonumber\\
\left\vert 13\right\rangle  &  =\mathbf{|-\mathbf{a}}_{1}\mathbf{,+\mathbf{a}%
}_{2}\mathbf{,+\mathbf{a}}_{3}\mathbf{\mathbf{,+a}}_{4}\mathbf{\rangle
},\left\vert 14\right\rangle =\mathbf{|-\mathbf{a}}_{1}\mathbf{,+\mathbf{a}%
}_{2}\mathbf{,-\mathbf{a}}_{3}\mathbf{\mathbf{,-a}}_{4}\mathbf{\rangle
},\nonumber\\
\left\vert 15\right\rangle  &  =|\mathbf{\mathbf{-a}}_{1}\mathbf{,-\mathbf{a}%
}_{2}\mathbf{,+\mathbf{a}}_{3}\mathbf{\mathbf{,-a}}_{4}\rangle,\left\vert
16\right\rangle =|\mathbf{\mathbf{-a}}_{1}\mathbf{,-\mathbf{a}}_{2}%
\mathbf{,-\mathbf{a}}_{3}\mathbf{\mathbf{,+a}}_{4}\rangle. \label{base2}%
\end{align}
They are the eigenstates of spin correlation operator
\[
\hat{\Omega}\left(  a_{1},a_{2},a_{3},a_{4}\right)  =\left(  \hat{s}%
\cdot\mathbf{a}_{1}\right)  \otimes\left(  \hat{s}\cdot\mathbf{a}_{2}\right)
\otimes\left(  \hat{s}\cdot\mathbf{a}_{3}\right)  \otimes\left(  \hat{s}%
\cdot\mathbf{a}_{4}\right)  .
\]
with the eigenvalues $\pm(1/2)^{4}$ for the states labeled respectively from
$1-8$ and $9-16$. The average of measuring outcome correlation from four
observers becomes the algebraic sum of the density operator. The local part of
correlation%
\begin{equation}
p_{lc}(a_{1},a_{2},a_{3},a_{4})=\sum_{i=1}^{8}\rho_{ii}^{lc}-\sum_{i=9}%
^{16}\rho_{ii}^{lc}=%
{\displaystyle\prod\limits_{i=1}^{4}}
\cos\theta_{a_{i}}, \label{4plc}%
\end{equation}
gives rise to the four-particle GBI such that%
\begin{align*}
&  p_{lc}\left(  a_{1},a_{2},a_{3},a_{4}\right)  p_{lc}\left(  a_{2}%
,a_{3},a_{4},a_{5}\right) \\
&  \times p_{lc}\left(  a_{3},a_{4},a_{5},a_{6}\right)  p_{lc}\left(
a_{4},a_{5},a_{6},a_{7}\right) \\
&  =\left(
{\displaystyle\prod\limits_{i=1}^{4}}
\cos^{2}\theta_{a_{2i-1}}\right)  \left(
{\displaystyle\prod\limits_{i=2}^{6}}
\cos^{2}\theta_{a_{i}}\right)  \cos^{2}\theta_{a_{4}}\\
&  \leq\left\vert p_{lc}\left(  a_{1},a_{3},a_{5},a_{7}\right)  \right\vert .
\end{align*}
It may be worthwhile to remark that the four-particle local part of
correlation Eq.(\ref{4plc}), which is independent of the state parameters
$\xi$, $\eta$, has a positive sign compared with the three-particle case
Eq.(\ref{plc}).

\subsubsection{Maximum violation}

The non-local elements of density operator are%
\[
\rho_{ii}^{nlc}=\pm\frac{1}{2^{4}}\sin\left(  2\xi\right)  \left(
{\displaystyle\prod\limits_{i=1}^{4}}
\sin\theta_{a_{i}}\right)  \cos\left(
{\displaystyle\sum\limits_{i=1}^{4}}
\phi_{a_{i}}+2\eta\right)  ,
\]
respectively for $i=1-8$ and $9-16$.

The non-local part of correlation is%
\[
p_{nlc}=\sin\left(  2\xi\right)  \left(
{\displaystyle\prod\limits_{i=1}^{4}}
\sin\theta_{a_{i}}\right)  \cos\left(
{\displaystyle\sum\limits_{i=1}^{4}}
\phi_{a_{i}}+2\eta\right)  .
\]
The four-particle quantum correlation-probability difference is defined by%
\begin{align*}
p_{GB}  &  =p\left(  a_{1},a_{2},a_{3},a_{4}\right)  p\left(  a_{2}%
,a_{3},a_{4},a_{5}\right)  p\left(  a_{3},a_{4},a_{5},a_{6}\right) \\
&  \times p\left(  a_{4},a_{5},a_{6},a_{7}\right)  -\left\vert p\left(
a_{1},a_{3},a_{5},a_{7}\right)  \right\vert ,
\end{align*}
any positive value of which indicates the violation of GBI. The maximum
violation appears when $\theta_{a_{1}}=\theta_{a_{2}}=\theta_{a_{3}}%
=\theta_{a_{4}}=\pi/2$, where we have%
\[
p(a_{1},a_{2},a_{3},a_{4})=\sin\left(  2\xi\right)  \cos\left(
{\displaystyle\sum\limits_{i=1}^{4}}
\phi_{a_{i}}+2\eta\right)  .
\]
Then the quantum correlation-probability difference becomes
\begin{align*}
p_{GB}  &  =\sin\left(
{\displaystyle\sum\limits_{i=1}^{4}}
\phi_{a_{i}}\right)  \sin\left(
{\displaystyle\sum\limits_{i=2}^{5}}
\phi_{a_{i}}\right)  \sin\left(
{\displaystyle\sum\limits_{i=3}^{6}}
\phi_{a_{i}}\right) \\
&  \times\sin\left(
{\displaystyle\sum\limits_{i=4}^{7}}
\phi_{a_{i}}\right)  -\left\vert \sin\left(
{\displaystyle\sum\limits_{i=0}^{3}}
\phi_{a_{2i+1}}\right)  \right\vert
\end{align*}
with parameters $\xi=\eta=\pi/4$. For the azimuthal angles $\phi_{a_{2i+1}}=0$
with $i=0,1,2,3$ we have%
\[
p_{GB}=\sin^{2}\left(  \phi_{a_{2}}+\phi_{a_{4}}\right)  \sin^{2}\left(
\phi_{a_{4}}+\phi_{a_{6}}\right)
\]
The maximum violation is%
\begin{equation}
p_{GB}^{\max}=1 \label{max1}%
\end{equation}
with the azimuthal angles $\phi_{a_{2i}}=\pi/4$ ($i=1,2,3$) or $\phi_{a_{4}%
}=0$, $\phi_{a_{2}}=\phi_{a_{6}}=\pi/2$.

\subsection{N-particle entangled state}

For $n$-particle entangled state
\[
\left\vert \psi\right\rangle =c_{1}|+\rangle^{\otimes n}+c_{2}|-\rangle
^{\otimes n},
\]
the GBI is satisfied by the local realistic mode with correlation%
\[
p_{lc}(a_{1},a_{2},...,a_{n})=\sum_{i=1}^{2^{n-1}}\rho_{ii}^{lc}%
-\sum_{i=2^{n-1}+1}^{2^{n}}\rho_{ii}^{lc},
\]
which gives rise to
\begin{equation}
p_{lc}(a_{1},a_{2},...,a_{n})=-\cos\left(  2\xi\right)
{\displaystyle\prod\limits_{i=1}^{n}}
\cos\theta_{a_{i}} \label{oddn}%
\end{equation}
for $n$ being odd number and
\begin{equation}
p_{lc}(a_{1},a_{2},...,a_{n})=%
{\displaystyle\prod\limits_{i=1}^{n}}
\cos\theta_{a_{i}} \label{evenn}%
\end{equation}
for even $n$.

\subsubsection{Odd n}

When $n$ is odd, the local part of normalized correlation probability
Eq.(\ref{oddn}) satisfies GBI that
\begin{align}
&  p_{lc}\left(  a_{1},a_{2},...,a_{n}\right)  p_{lc}\left(  a_{2}%
,a_{3},...,a_{n+1}\right) \nonumber\\
&  \times p_{lc}\left(  a_{3},a_{4},...,a_{n+2}\right)  ...p_{lc}\left(
a_{n},a_{n+1},...,a_{2n-1}\right) \nonumber\\
&  =\cos^{n-1}\left(  2\xi\right)
{\displaystyle\prod\limits_{i=2}^{n+1}}
\cos^{2}\theta_{a_{i}}%
{\displaystyle\prod\limits_{i=4}^{n+3}}
\cos^{2}\theta_{a_{i}}%
{\displaystyle\prod\limits_{i=6}^{n+5}}
\cos^{2}\theta_{a_{i}}\nonumber\\
&  \times...\times%
{\displaystyle\prod\limits_{i=n-1}^{2n-2}}
\cos^{2}\theta_{a_{i}}p_{lc}\left(  a_{1},a_{3},...,a_{2n-1}\right)
\nonumber\\
&  \leq\left\vert p_{lc}\left(  a_{1},a_{3},...,a_{2n-1}\right)  \right\vert .
\label{oddGBI}%
\end{align}

With the non-local part of correlation
\begin{align}
&  p_{nlc}(a_{1},a_{2},...,a_{n})\nonumber\\
&  =\sin\left(  2\xi\right)  \left(
{\displaystyle\prod\limits_{i=1}^{n}}
\sin\theta_{a_{i}}\right)  \cos\left(
{\displaystyle\sum\limits_{i=1}^{n}}
\phi_{a_{i}}+2\eta\right)  , \label{oddpnlc}%
\end{align}
the total correlation becomes%
\begin{align*}
&  p(a_{1},a_{2},...,a_{n})\\
&  =-\cos\left(  2\xi\right)
{\displaystyle\prod\limits_{i=1}^{n}}
\cos\theta_{a_{i}}\\
&  +\sin\left(  2\xi\right)  \left(
{\displaystyle\prod\limits_{i=1}^{n}}
\sin\theta_{a_{i}}\right)  \cos\left(
{\displaystyle\sum\limits_{i=1}^{n}}
\phi_{a_{i}}+2\eta\right)  .
\end{align*}
For the polar angles $\theta_{a_{i}}=\pi/2$, and state parameters $\xi
=\pi/4\mathrm{mod}2\pi\ $and $\eta=\pi/4\mathrm{mod}2\pi$, it is simplified as%
\[
p(a_{1},a_{2},...,a_{n})=-\sin\left(
{\displaystyle\sum\limits_{i=1}^{n}}
\phi_{a_{i}}\right)  .
\]
Then the quantum correlation-probability difference becomes%
\begin{align*}
p_{GB}  &  =p\left(  a_{1},a_{2},...,a_{n}\right)  p\left(  a_{2}%
,a_{3},...,a_{n+1}\right)  p\left(  a_{3},a_{4},...,a_{n+2}\right) \\
&  \times...\times p\left(  a_{n},a_{n+1},...,a_{2n-1}\right)  -\left\vert
p\left(  a_{1},a_{3},...,a_{2n-1}\right)  \right\vert \\
&  =-\sin\left(
{\displaystyle\sum\limits_{i=1}^{n}}
\phi_{a_{i}}\right)  \sin\left(
{\displaystyle\sum\limits_{i=2}^{n+1}}
\phi_{a_{i}}\right)  \times...\\
&  \times\sin\left(
{\displaystyle\sum\limits_{i=n}^{2n-1}}
\phi_{a_{i}}\right)  -\left\vert \sin\left(
{\displaystyle\sum\limits_{i=1}^{n}}
\phi_{a_{2i-1}}\right)  \right\vert
\end{align*}
When $\phi_{a_{2i-1}}=0$ ($i=1,2,...n$)$,$ we have%
\begin{align}
p_{GB}  &  =-\sin\left(
{\displaystyle\sum\limits_{i=1}^{\left(  n-1\right)  /2}}
\phi_{a_{2i}}\right)  \sin\left(
{\displaystyle\sum\limits_{i=1}^{\left(  n+1\right)  /2}}
\phi_{a_{2i}}\right) \nonumber\\
&  \times\sin\left(
{\displaystyle\sum\limits_{i=2<n-1}^{\left(  n+1\right)  /2}}
\phi_{a_{2i}}\right)  \sin\left(
{\displaystyle\sum\limits_{i=2<n-1}^{\left(  n+3\right)  /2}}
\phi_{a_{2i}}\right) \nonumber\\
&  \times\sin\left(
{\displaystyle\sum\limits_{i=3<n-1}^{\left(  n+3\right)  /2}}
\phi_{a_{2i}}\right)  \sin\left(
{\displaystyle\sum\limits_{i=3<n-1}^{\left(  n+5\right)  /2}}
\phi_{a_{2i}}\right) \nonumber\\
&  \times...\times\sin\left(
{\displaystyle\sum\limits_{i=\left(  n+1\right)  /2}^{n-1}}
\phi_{a_{2i}}\right)  , \label{pgb}%
\end{align}
which possesses a maximum value with the azimuthal angles given by
$\phi_{a_{2i}}=0$ except $i=(n\pm1)/2$, and $\phi_{a_{n-1}}=\phi_{a_{n+1}%
}=3\pi/4$. Along these measuring conditions the correlation-probability
difference Eq.(\ref{pgb}) approaches the maximum bound%
\begin{align}
p_{GB}^{\max}  &  =-\sin\phi_{a_{n-1}}\sin\phi_{a_{n+1}}\sin^{n-2}\left(
\phi_{a_{n-1}}+\phi_{a_{n+1}}\right) \nonumber\\
&  =\frac{1}{2}. \label{gbimax}%
\end{align}

\subsubsection{Even n}

For even $n$, the local part of $n$-particle correlation probability satisfies
the GBI that
\begin{align}
&  p_{lc}\left(  a_{1},a_{2},...,a_{n}\right)  p_{lc}\left(  a_{2}%
,a_{3},...,a_{n+1}\right) \nonumber\\
&  \times p_{lc}\left(  a_{3},a_{4},...,a_{n+2}\right)  ...p_{lc}\left(
a_{n},a_{n+1},...,a_{2n-1}\right) \nonumber\\
&  =%
{\displaystyle\prod\limits_{i=1}^{n}}
\cos\theta_{a_{i}}%
{\displaystyle\prod\limits_{i=2}^{n+1}}
\cos\theta_{a_{i}}%
{\displaystyle\prod\limits_{i=3}^{n+2}}
\cos\theta_{a_{i}}...%
{\displaystyle\prod\limits_{i=n}^{2n-1}}
\cos\theta_{a_{i}}\nonumber\\
&  =%
{\displaystyle\prod\limits_{i=2}^{n}}
\cos^{2}\theta_{a_{i}}%
{\displaystyle\prod\limits_{i=4}^{n+2}}
\cos^{2}\theta_{a_{i}}%
{\displaystyle\prod\limits_{i=6}^{n+4}}
\cos^{2}\theta_{a_{i}}...%
{\displaystyle\prod\limits_{i=n}^{2n-2}}
\cos^{2}\theta_{a_{i}}\nonumber\\
&  \times p_{lc}\left(  a_{1},a_{3},...,a_{2n-1}\right) \nonumber\\
&  \leq\left\vert p_{lc}\left(  a_{1},a_{3},...,a_{2n-1}\right)  \right\vert .
\label{egbi}%
\end{align}
The non-local part of correlation $p_{nlc}(a_{1},a_{2},...,a_{n})$ is the same
as that of odd $n$ given in Eq.(\ref{oddpnlc}).

The normalized total correlation
\begin{align*}
&  p(a_{1},a_{2},...,a_{n})\\
&  =%
{\displaystyle\prod\limits_{i=1}^{n}}
\cos\theta_{a_{i}}+\sin\left(  2\xi\right) \\
&  \times\left(
{\displaystyle\prod\limits_{i=1}^{n}}
\sin\theta_{a_{i}}\right)  \cos\left(
{\displaystyle\sum\limits_{i=1}^{n}}
\phi_{a_{i}}+2\eta\right)  ,
\end{align*}
is simplified as%
\[
p(a_{1},a_{2},...,a_{n})=-\sin\left(
{\displaystyle\sum\limits_{i=1}^{n}}
\phi_{a_{i}}\right)
\]
with the state parameters $\xi=\eta=\pi/4\mathrm{mod}2\pi$, and polar angles
$\theta_{a_{i}}=\pi/2$ of measuring directions. Then the correlation
difference becomes%
\begin{align*}
p_{GB}  &  =\sin\left(
{\displaystyle\sum\limits_{i=1}^{n}}
\phi_{a_{i}}\right)  \sin\left(
{\displaystyle\sum\limits_{i=2}^{n+1}}
\phi_{a_{i}}\right)  \sin\left(
{\displaystyle\sum\limits_{i=3}^{n+2}}
\phi_{a_{i}}\right) \\
&  \times...\times\sin\left(
{\displaystyle\sum\limits_{i=n}^{2n-1}}
\phi_{a_{i}}\right)  -\left\vert \sin\left(
{\displaystyle\sum\limits_{i=1}^{n}}
\phi_{a_{2i-1}}\right)  \right\vert .
\end{align*}
The maximum value of $p_{GB}$ appears when $\phi_{a_{i}}=0,$ for
$i=1,3,5,...,2n-1,$ we have%
\begin{align*}
p_{GB}  &  =\sin^{2}\left(
{\displaystyle\sum\limits_{i=1}^{n/2}}
\phi_{a_{2i}}\right)  \sin^{2}\left(
{\displaystyle\sum\limits_{i=2}^{n/2+1}}
\phi_{a_{2i}}\right) \\
&  \times\sin^{2}\left(
{\displaystyle\sum\limits_{i=3}^{n/2+2}}
\phi_{a_{2i}}\right)  ...\sin^{2}\left(
{\displaystyle\sum\limits_{i=n/2}^{n-1}}
\phi_{a_{2i}}\right)  .
\end{align*}
The maximum bound of violation is%
\begin{equation}
p_{GB}^{\max}=1, \label{maxeven}%
\end{equation}
under the condition of $\phi_{a_{2i}}=\pi/n$ with $i=1,2,3,...,n-1$. For the
$n$-particle entangled state of spin-$1/2$ the GBI is always satisfied by the
local correlation. The non-local part of correlation gives rise to the
violation of GBI. The maximal violation bound is $p_{GB}^{\max}=1/2$ for the
odd $n$ and $1$ for the even $n$. The maximum violation takes place when the
$2n-1$ measuring-directions are perpendicular to the spin polarization
($z$-axis), the maximum bound depends on the $\sin$ function of $n/2$
azimuthal angles $\phi_{a_{2i}}$ for the even $n$. We always have the
possibility to choose the equal value of the angles $\phi_{a_{2i}}=\pi/n$ to
approach the maximum bound $p_{GB}^{\max}=1$, which is in agreement with the
previous observation, that the violation is larger\cite{ex4} for even $n$.

\section{Spin parity effect in the violation of GBI for $n$-particle entangled
Schr\"{o}dinger-cat-state of spin-$s$}

For entangled Schr\"{o}dinger-cat-states of spin-$s$, the GBI is always
satisfied since the non-local correlation vanishes by quantum average. We now
consider the measuring outcomes restricted in the subspace of SCS instead,
namely, only the maximum spin values $\pm s$ are measured along arbitrary directions.

\subsection{Three-particle case}

The local and non-local parts of density operator $\hat{\rho}$ are
respectively written as
\begin{align*}
\hat{\rho}_{lc}  &  =\sin^{2}\xi|+s,+s,+s\rangle\left\langle
+s,+s,+s\right\vert \\
&  +\cos^{2}\xi|-s,-s,-s\rangle\left\langle -s,-s,-s\right\vert ,\\
\hat{\rho}_{nlc}  &  =\sin\xi\cos\xi\left(
\begin{array}
[c]{c}%
e^{i2\eta}|+s,+s,+s\rangle\left\langle -s,-s,-s\right\vert \\
+e^{-i2\eta}|-s,-s,-s\rangle\left\langle +s,+s,+s\right\vert
\end{array}
\right)  .
\end{align*}
for the three-particle entangled cat state%
\[
\left\vert \psi\right\rangle _{GHZ}=c_{1}|+s,+s,+s\rangle+c_{2}%
|-s,-s,-s\rangle.
\]
The SCSs of projection spin-operator in direction $\mathbf{r}=\mathbf{a}%
_{1},\mathbf{a}_{2},\mathbf{a}_{3}$ are found from the eigenequations
\[
\hat{s}\cdot\mathbf{r}\left\vert \pm\mathbf{r}\right\rangle =\mathbf{\pm
}s\left\vert \pm\mathbf{r}\right\rangle ,
\]
as\cite{p1,p2,p5,coherent1,coherent2,Dicke}%
\begin{align*}
\left\vert +\mathbf{r}\right\rangle  &  =%
{\textstyle\sum\limits_{m=-s}^{s}}
\left(
\begin{array}
[c]{l}%
2s\\
s+m
\end{array}
\right)  ^{\frac{1}{2}}K_{r}^{s+m}\Gamma_{r}^{s-m}\exp\left[  i\left(
s-m\right)  \phi_{r}\right]  \left\vert m\right\rangle ,\\
\left\vert -\mathbf{r}\right\rangle  &  =%
{\textstyle\sum\limits_{m=-s}^{s}}
\left(
\begin{array}
[c]{l}%
2s\\
s+m
\end{array}
\right)  ^{\frac{1}{2}}K_{r}^{s-m}\Gamma_{r}^{s+m}\\
&  \times\exp\left[  i\left(  s-m\right)  \left(  \phi_{r}+\pi\right)
\right]  \left\vert m\right\rangle .
\end{align*}
With the independent measuring-outcome basis vectors labeled in Eq.(\ref{base}%
) the local part of measuring outcome correlation is%
\[
P_{lc}(a_{1},a_{2},a_{3})=s^{3}\left(  \sum_{i=1}^{4}\rho_{ii}^{lc}-\sum
_{i=5}^{8}\rho_{ii}^{lc}\right)  ,
\]
in which the elements of local density operator are given by the same formulas
as Eq.(\ref{element}), however, with the power index $"2"$ replaced by $"4s"$,
for example the first one is
\[
\rho_{11}^{lc}=\sin^{2}\xi%
{\displaystyle\prod\limits_{i=1}^{3}}
K_{a_{i}}^{4s}+\cos^{2}\xi%
{\displaystyle\prod\limits_{i=1}^{3}}
\Gamma_{a_{i}}^{4s}.
\]
The normalized local correlation is found as%
\begin{align*}
&  p_{lc}(a_{1},a_{2},a_{3})\\
&  =\frac{P_{lc}}{s^{3}}\\
&  =-\cos\left(  2\xi\right)  \left(  K_{a_{1}}^{4s}-\Gamma_{a_{1}}%
^{4s}\right)  \left(  K_{a_{2}}^{4s}-\Gamma_{a_{2}}^{4s}\right)  \left(
K_{a_{3}}^{4s}-\Gamma_{a_{3}}^{4s}\right)  ,
\end{align*}
which gives rise to the GBI such that%
\begin{align*}
&  p_{lc}\left(  a_{1},a_{2},a_{3}\right)  p_{lc}\left(  a_{2},a_{3}%
,a_{4}\right)  p_{lc}\left(  a_{3},a_{4},a_{5}\right) \\
&  \leq-\cos\left(  2\xi\right)  \left(  K_{a_{1}}^{4s}-\Gamma_{a_{1}}%
^{4s}\right)  \left(  K_{a_{3}}^{4s}-\Gamma_{a_{3}}^{4s}\right)  \left(
K_{a_{5}}^{4s}-\Gamma_{a_{5}}^{4s}\right) \\
&  \leq\left\vert p_{lc}\left(  a_{1},a_{3},a_{5}\right)  \right\vert .
\end{align*}
The non-local elements of density operator are seen to be%
\begin{align*}
\rho_{11}^{nlc}  &  =\sin\left(  2\xi\right)  K_{a_{1}}^{2s}\Gamma_{a_{1}%
}^{2s}K_{a_{2}}^{2s}\Gamma_{a_{2}}^{2s}K_{a_{3}}^{2s}\Gamma_{a_{3}}^{2s}\\
&  \times\cos\left[  2s\left(  \phi_{a_{1}}+\phi_{a_{2}}+\phi_{a_{3}}\right)
+2\eta\right] \\
&  =\rho_{ii}^{nlc},
\end{align*}
for $i=2-4$ and%
\[
\rho_{jj}^{nlc}=\left(  -1\right)  ^{2s}\rho_{11}^{nlc}%
\]
with $j=5-8$. It may be worthwhile to notice that the density matrix elements
of non-local part differ by a phase factor $(-1)^{2s}=\exp(i2s\pi)$, which
resulted from the geometric phase between SCSs of the north- and south- pole
gauges. For integer spin-$s$, the non-local correlation vanishes,
\[
p_{nlc}(a_{1},a_{2},a_{3})=\sum_{i=1}^{4}\rho_{ii}^{nlc}-\sum_{i=5}^{8}%
\rho_{ii}^{nlc},
\]
which leads to non-violation of GBI. For half-integer spin-$s$, the non-local
correlation becomes%
\begin{align*}
p_{nlc}(a_{1},a_{2},a_{3})  &  =2^{-3(2s-1)}\sin\left(  2\xi\right)  \left(
{\displaystyle\prod\limits_{i=1}^{3}}
\sin^{2s}\theta_{a_{i}}\right) \\
&  \times\cos\left[  2s\left(
{\displaystyle\sum\limits_{i=1}^{3}}
\phi_{a_{i}}\right)  +2\eta\right]  .
\end{align*}
The whole quantum correlation-probability $p(a_{1},a_{2},a_{3})$ can approach
the maximum violation bound with polar angle $\theta_{r}=\pi/2$ that%
\begin{equation}
p(a_{1},a_{2},a_{3})=2^{-3(2s-1)}\sin\left(  2\xi\right)  \cos\left[
2s\left(
{\displaystyle\sum\limits_{i=1}^{3}}
\phi_{a_{i}}\right)  +2\eta\right]  .
\end{equation}
The correlation probability for the measurement in the SCS subspace decreases
with the increase of spin $s$, since the dimension of whole Hilbert space is
$(2s+1)^{3}$ , while the number of measuring outcome states is only
$8$.\textbf{ }The correlation probability vanishes when $s\rightarrow\infty$
in agreement with the known observations\cite{o1,o2,o3,o4}. We may consider
the relative or scaled correlation probability%
\[
p_{rl}\left(  a_{1},a_{2},a_{3}\right)  =\frac{p\left(  a_{1},a_{2}%
,a_{3}\right)  }{N}%
\]
where normalization constant%
\[
N=%
{\displaystyle\sum\limits_{i=1}^{8}}
|\langle i|\psi\rangle|^{2}=%
{\displaystyle\sum\limits_{i=1}^{8}}
\rho_{ii}=2^{-3(2s-1)},
\]
is the total probability of entangled state $|\psi\rangle$ in the eight
measuring basis vectors of SCS given by Eq.(\ref{base}).

The relative or scaled correlation probability is%
\begin{equation}
p_{rl}\left(  a_{1},a_{2},a_{3}\right)  =\sin\left(  2\xi\right)  \cos\left[
2s\left(
{\displaystyle\sum\limits_{i=1}^{3}}
\phi_{a_{i}}\right)  +2\eta\right]  \label{rl}%
\end{equation}
In the following the scaled correlation probabilities of Eq.(\ref{rl}) is
adopted without the subscript "$rl$" for the sake of simplicity.

The quantity of correlation difference is found as
\[
p_{GB}=-\sin\left(  2s\phi_{a_{2}}\right)  \sin\left[  2s\left(  \phi_{a_{2}%
}+\phi_{a_{4}}\right)  \right]  \sin\left(  2s\phi_{a_{4}}\right)
\]
with parameters $\xi=\eta=\pi/4$, and azimuthal angles of measuring
directions$\ \phi_{a_{1}}=\phi_{a_{3}}=\phi_{a_{5}}=0$. The maximum bound of
violation is%
\begin{equation}
p_{GB}^{\max}=\frac{1}{2} \label{max3s}%
\end{equation}
when $\phi_{a_{2}}=\phi_{a_{4}}=3\pi/\left(  8s\right)  $ in consistence with
the case of spin-$1/2$.

\subsection{Four-particle case}

For four-particle entangled Schr\"{o}dinger-cat-state%
\[
\left\vert \psi\right\rangle _{GHZ}=c_{1}|+s,+s,+s,+s\rangle+c_{2}%
|-s,-s,-s,-s\rangle,
\]
\ we have $16$ independent basis vectors for the measuring along $4$ arbitrary
directions $\mathbf{a}_{1},\mathbf{a}_{2},\mathbf{a}_{3}\mathbf{,a}_{4}$.
Following the same procedure we have the local correlation probability%
\[
p_{lc}(a_{1},a_{2},a_{3},a_{4})=%
{\displaystyle\prod\limits_{i=1}^{4}}
\left(  K_{a_{i}}^{4s}-\Gamma_{a_{i}}^{4s}\right)  ,
\]
which gives rise to the GBI that%
\begin{align*}
&  p_{lc}\left(  a_{1},a_{2},a_{3},a_{4}\right)  p_{lc}\left(  a_{2}%
,a_{3},a_{4},a_{5}\right) \\
&  \times p_{lc}\left(  a_{3},a_{4},a_{5},a_{6}\right)  p_{lc}\left(
a_{4},a_{5},a_{6},a_{7}\right) \\
&  \leq%
{\displaystyle\prod\limits_{i=0}^{3}}
\left(  K_{a_{2i+1}}^{4s}-\Gamma_{a_{2i+1}}^{4s}\right) \\
&  \leq\left\vert p_{lc}\left(  a_{1},a_{3},a_{5},a_{7}\right)  \right\vert .
\end{align*}

The non-local correlation vanishes for integer spin-$s$. For half-integer $s$,
it is
\begin{align*}
p_{nlc}(a_{1},a_{2},a_{3},a_{4})  &  =2^{-4(2s-1)}\sin\left(  2\xi\right)
\left(
{\displaystyle\prod\limits_{i=1}^{4}}
\sin^{2s}\theta_{a_{i}}\right) \\
&  \times\cos\left[  2s\left(
{\displaystyle\sum\limits_{i=1}^{4}}
\phi_{a_{i}}\right)  +2\eta\right]  .
\end{align*}
The total quantum correlation-probability is%
\begin{align*}
p(a_{1},a_{2},a_{3},a_{4})  &  =2^{-4(2s-1)}\sin\left(  2\xi\right) \\
&  \times\cos\left[  2s\left(
{\displaystyle\sum\limits_{i=1}^{4}}
\phi_{a_{i}}\right)  +2\eta\right]  ,
\end{align*}
under the condition of polar angles equal to $\theta_{i}=\pi/2$. We again
consider the relative or scaled correlation probability $p_{rl}\left(
a_{1},a_{2},a_{3},a_{4}\right)  =p\left(  a_{1},a_{2},a_{3},a_{4}\right)  /N$
( $N=2^{-4(2s-1)}$), which becomes%
\[
p_{rl}(a_{1},a_{2},a_{3},a_{4})=\sin\left(  2\xi\right)  \cos\left[  2s\left(
%
{\displaystyle\sum\limits_{i=1}^{4}}
\phi_{a_{i}}\right)  +2\eta\right]  .
\]
Then the quantum correlation-probability difference is%
\begin{align*}
p_{GB}  &  =p\left(  a_{1},a_{2},a_{3},a_{4}\right)  p\left(  a_{2}%
,a_{3},a_{4},a_{5}\right)  p\left(  a_{3},a_{4},a_{5},a_{6}\right) \\
&  \times p\left(  a_{4},a_{5},a_{6},a_{7}\right)  -\left\vert p_{lc}\left(
a_{1},a_{3},a_{5},a_{7}\right)  \right\vert \\
&  =\sin^{2}\left[  2s\left(  \phi_{a_{2}}+\phi_{a_{4}}\right)  \right]
\sin^{2}\left[  2s\left(  \phi_{a_{4}}+\phi_{a_{6}}\right)  \right]
\end{align*}
with parameters$\ \xi=\eta=\pi/4$, and azimuthal angles$\ \phi_{a_{1}}%
=\phi_{a_{3}}=\phi_{a_{5}}=\phi_{a_{7}}=0$. The maximum violation bound is
obviously%
\[
p_{GB}^{\max}=1
\]
when $\phi_{a_{2}}=\phi_{a_{4}}=\phi_{a_{6}}=\pi/\left(  8s\right)  $.

\subsection{N-particle case}

For $n$-particle state
\[
\left\vert \psi\right\rangle =e^{i\eta}\sin\xi|+s\rangle^{\otimes n}%
+e^{-i\eta}\cos\xi|-s\rangle^{\otimes n},
\]
the local part of $n$-direction measuring outcome correlation%
\[
p_{lc}(a_{1},a_{2},...,a_{n})=\sum_{i=1}^{2^{n-1}}\rho_{ii}^{lc}%
-\sum_{i=2^{n-1}+1}^{2^{n}}\rho_{ii}^{lc},
\]
is found respectively as%
\begin{equation}
p_{lc}(a_{1},a_{2},...,a_{n})=-\cos\left(  2\xi\right)
{\displaystyle\prod\limits_{i=1}^{n}}
\left(  K_{a_{i}}^{4s}-\Gamma_{a_{i}}^{4s}\right)  \label{oddns}%
\end{equation}
for odd $n$ and%
\begin{equation}
p_{lc}(a_{1},a_{2},...,a_{n})=%
{\displaystyle\prod\limits_{i=1}^{n}}
\left(  K_{a_{i}}^{4s}-\Gamma_{a_{i}}^{4s}\right)  \label{evenns}%
\end{equation}
for even $n$. The expressions of Eqs.(\ref{oddns},\ref{evenns}) have the same
forms compared with $n$-particle state of spin-$1/2$ local correlation
Eq.(\ref{oddn}) and Eq.(\ref{evenn}), where $\cos\theta_{a_{i}}$ is replaced
by $\left(  K_{a_{i}}^{4s}-\Gamma_{a_{i}}^{4s}\right)  $. The GBI is satisfied
by the $n$-particle local correlations that%
\begin{align*}
&  p_{lc}\left(  a_{1},a_{2},...,a_{n}\right)  p_{lc}\left(  a_{2}%
,a_{3},...,a_{n+1}\right)  ...p_{lc}\left(  a_{n},a_{n+1},...,a_{2n-1}\right)
\\
&  =\cos^{n-1}\left(  2\xi\right)
{\displaystyle\prod\limits_{i=2}^{n+1}}
\left(  K_{a_{i}}^{4s}-\Gamma_{a_{i}}^{4s}\right)  ^{2}%
{\displaystyle\prod\limits_{i=4}^{n+3}}
\left(  K_{a_{i}}^{4s}-\Gamma_{a_{i}}^{4s}\right)  ^{2}\\
&  \times%
{\displaystyle\prod\limits_{i=6}^{n+5}}
\left(  K_{a_{i}}^{4s}-\Gamma_{a_{i}}^{4s}\right)  ^{2}...%
{\displaystyle\prod\limits_{i=n-1}^{2n-2}}
\left(  K_{a_{i}}^{4s}-\Gamma_{a_{i}}^{4s}\right)  ^{2}p_{lc}\left(
a_{1},a_{3},...,a_{2n-1}\right) \\
&  \leq\left\vert p_{lc}\left(  a_{1},a_{3},...,a_{2n-1}\right)  \right\vert .
\end{align*}
with odd $n$, and%
\begin{align*}
&  p_{lc}\left(  a_{1},a_{2},...,a_{n}\right)  p_{lc}\left(  a_{2}%
,a_{3},...,a_{n+1}\right)  ...p_{lc}\left(  a_{n},a_{n+1},...,a_{2n-1}\right)
\\
&  =%
{\displaystyle\prod\limits_{i=2}^{n}}
\left(  K_{a_{i}}^{4s}-\Gamma_{a_{i}}^{4s}\right)  ^{2}%
{\displaystyle\prod\limits_{i=4}^{n+2}}
\left(  K_{a_{i}}^{4s}-\Gamma_{a_{i}}^{4s}\right)  ^{2}%
{\displaystyle\prod\limits_{i=6}^{n+4}}
\left(  K_{a_{i}}^{4s}-\Gamma_{a_{i}}^{4s}\right)  ^{2}\\
&  \times...\times%
{\displaystyle\prod\limits_{i=n}^{2n-2}}
\left(  K_{a_{i}}^{4s}-\Gamma_{a_{i}}^{4s}\right)  ^{2}p_{lc}\left(
a_{1},a_{3},...,a_{2n-1}\right) \\
&  \leq\left\vert p_{lc}\left(  a_{1},a_{3},...,a_{2n-1}\right)  \right\vert .
\end{align*}
with even $n$. In the following we find the maximum violation bound including
the non-local correlations. Since the non-local correlation of integer
spin-$s$ vanishes, the violation of GBI appears only for the half-integer
spin-$s$.

\subsubsection{Maximum violation bound for odd $n$\ }

Including the non-local part of correlation
\begin{align*}
p_{nlc}(a_{1},a_{2},...,a_{n})  &  =2^{-n(2s-1)}\sin\left(  2\xi\right) \\
&  \times\left(
{\displaystyle\prod\limits_{i=1}^{n}}
\sin^{2s}\theta_{a_{i}}\right)  \cos\left[  2s\left(
{\displaystyle\sum\limits_{i=1}^{n}}
\phi_{a_{i}}\right)  +2\eta\right]  ,
\end{align*}
the quantum correlation is%
\begin{align*}
p(a_{1},a_{2},...,a_{n})  &  =-\cos\left(  2\xi\right)
{\displaystyle\prod\limits_{i=1}^{n}}
\left(  K_{a_{i}}^{4s}-\Gamma_{a_{i}}^{4s}\right) \\
&  +2^{-n(2s-1)}\sin\left(  2\xi\right)  \left(
{\displaystyle\prod\limits_{i=1}^{n}}
\sin^{2s}\theta_{a_{i}}\right) \\
&  \times\cos\left[  2s\left(
{\displaystyle\sum\limits_{i=1}^{n}}
\phi_{a_{i}}\right)  +2\eta\right]  .
\end{align*}
With polar angle $\theta_{r}=\pi/2$, where the maximum violation appears, we
have the simplified quantum correlation
\begin{align*}
p(a_{1},a_{2},...,a_{n})  &  =2^{-n(2s-1)}\sin\left(  2\xi\right) \\
&  \times\cos\left[  2s\left(
{\displaystyle\sum\limits_{i=1}^{n}}
\phi_{a_{i}}\right)  +2\eta\right]  .
\end{align*}
The decreasing number factor with $s$ can be scaled out by $p_{rl}\left(
a_{1},a_{2},...,a_{n}\right)  =p\left(  a_{1},a_{2},...,a_{n}\right)  /N$ with
$N=%
{\displaystyle\sum\limits_{i=1}^{2^{n}}}
|\langle i|\psi\rangle|^{2}=%
{\displaystyle\sum\limits_{i=1}^{2^{n}}}
\rho_{ii}=2^{-n(2s-1)}.$ The relative or scaled correlation probability
becomes%
\begin{equation}
p_{rl}\left(  a_{1},a_{2},...,a_{n}\right)  =\sin\left(  2\xi\right)
\cos\left[  2s\left(
{\displaystyle\sum\limits_{i=1}^{n}}
\phi_{a_{i}}\right)  +2\eta\right]  . \label{oddrl}%
\end{equation}
In the following the scaled correlation probabilities of Eq.(\ref{oddrl}) is
adopted without the subscript "$rl$" for the sake of simplicity.

With state parameters $\xi=\pi/4\mathrm{mod}2\pi\ $and $\eta=\pi
/4\mathrm{{mod}}2\pi$, Eq.(\ref{oddrl}) becomes%
\[
p\left(  a_{1},a_{2},...,a_{n}\right)  =-\sin\left(  2s%
{\displaystyle\sum\limits_{i=1}^{n}}
\phi_{a_{i}}\right)  .
\]
The quantum correlation-probability difference for the spin-$s$ case is seen
to be%
\begin{align*}
p_{GB}  &  =-\sin\left(  2s%
{\displaystyle\sum\limits_{i=1}^{n}}
\phi_{a_{i}}\right)  \sin\left(  2s%
{\displaystyle\sum\limits_{i=2}^{n+1}}
\phi_{a_{i}}\right) \\
&  \times...\times\sin\left(  2s%
{\displaystyle\sum\limits_{i=n}^{2n-1}}
\phi_{a_{i}}\right)  -\left\vert \sin\left(  2s%
{\displaystyle\sum\limits_{i=1,3,5}^{2n-1}}
\phi_{a_{i}}\right)  \right\vert ,
\end{align*}
which reduces to%
\begin{align*}
p_{GB}  &  =-\sin\left(  2s%
{\displaystyle\sum\limits_{i=1}^{\left(  n-1\right)  /2}}
\phi_{a_{2i}}\right)  \sin\left(  2s%
{\displaystyle\sum\limits_{i=1}^{\left(  n+1\right)  /2}}
\phi_{a_{2i}}\right) \\
&  \times\sin\left(  2s%
{\displaystyle\sum\limits_{i=2}^{\left(  n+1\right)  /2}}
\phi_{a_{2i}}\right)  \sin\left(  2s%
{\displaystyle\sum\limits_{i=2}^{\left(  n+3\right)  /2}}
\phi_{a_{2i}}\right) \\
&  \times\sin\left(  2s%
{\displaystyle\sum\limits_{i=3}^{\left(  n+3\right)  /2}}
\phi_{a_{2i}}\right)  \sin\left(  2s%
{\displaystyle\sum\limits_{i=3}^{\left(  n+5\right)  /2}}
\phi_{a_{2i}}\right) \\
\times &  ...\times\sin\left(  2s%
{\displaystyle\sum\limits_{i=\left(  n+1\right)  /2}^{n-1}}
\phi_{a_{2i}}\right)  .
\end{align*}
with vanishing angles $\phi_{a_{i}}=0,\left(  i=1,3,5,...,2n-1\right)  $ of
measuring directions. We, furthermore, let all other angles be zero that
$\phi_{a_{2i}}=0$ except the two angles $\phi_{a_{n-1}}=\phi_{a_{n+1}}%
=3\pi/\left(  8s\right)  $, then the maximum violation bound is approached%
\begin{align}
p_{GB}^{\max}  &  =-\sin\left(  2s\phi_{a_{n-1}}\right)  \sin\left(
2s\phi_{a_{n+1}}\right) \nonumber\\
&  \times\sin^{n-2}\left[  2s\left(  \phi_{a_{n-1}}+\phi_{a_{n+1}}\right)
\right] \nonumber\\
&  =\frac{1}{2}. \label{maxoddn}%
\end{align}

\subsubsection{Even n}

For even $n$ the quantum correlation including the non-local part is%
\begin{align*}
p(a_{1},a_{2},...,a_{n})  &  =%
{\displaystyle\prod\limits_{i=1}^{n}}
\left(  K_{a_{i}}^{4s}-\Gamma_{a_{i}}^{4s}\right)  +2^{-n(2s-1)}\\
&  \times\sin\left(  2\xi\right)  \left(
{\displaystyle\prod\limits_{i=1}^{n}}
\sin^{2s}\theta_{a_{i}}\right) \\
&  \times\cos\left[  2s\left(
{\displaystyle\sum\limits_{i=1}^{n}}
\phi_{a_{i}}\right)  +2\eta\right]  ,
\end{align*}
which reduces (after re-scale) to
\begin{equation}
p\left(  a_{1},a_{2},...,a_{n}\right)  =\sin\left(  2\xi\right)  \cos\left[
2s\left(
{\displaystyle\sum\limits_{i=1}^{n}}
\phi_{a_{i}}\right)  +2\eta\right]  , \label{evenrl}%
\end{equation}
with the polar angle $\theta_{r}=\pi/2$. For the state parameters $\xi
=\eta=\pi/4\mathrm{mod}2\pi$, the Eq.(\ref{evenrl}) becomes%
\[
p\left(  a_{1},a_{2},...,a_{n}\right)  =-\sin\left(  2s%
{\displaystyle\sum\limits_{i=1}^{n}}
\phi_{a_{i}}\right)  ,
\]
with which the quantum correlation-probability difference is%
\begin{align*}
p_{GB}  &  =\sin\left(  2s%
{\displaystyle\sum\limits_{i=1}^{n}}
\phi_{a_{i}}\right)  \sin\left(  2s%
{\displaystyle\sum\limits_{i=2}^{n+1}}
\phi_{a_{i}}\right)  \sin\left(  2s%
{\displaystyle\sum\limits_{i=3}^{n+2}}
\phi_{a_{i}}\right) \\
&  \times...\times\sin\left(  2s%
{\displaystyle\sum\limits_{i=n}^{2n-1}}
\phi_{a_{i}}\right)  -\left\vert \sin\left(  2s%
{\displaystyle\sum\limits_{i=1,3,5}^{2n-1}}
\phi_{a_{i}}\right)  \right\vert .
\end{align*}
Choosing azimuthal angles $\phi_{a_{i}}=0,\left(  i=1,3,5,...,2n-1\right)  $
of measuring directions we have%
\begin{align*}
p_{GB}  &  =\sin^{2}\left(  2s%
{\displaystyle\sum\limits_{i=1}^{n/2}}
\phi_{a_{2i}}\right)  \sin^{2}\left(  2s%
{\displaystyle\sum\limits_{i=2}^{n/2+1}}
\phi_{a_{2i}}\right) \\
&  \times\sin^{2}\left(  2s%
{\displaystyle\sum\limits_{i=3}^{n/2+2}}
\phi_{a_{2i}}\right)  ...\sin^{2}\left(  2s%
{\displaystyle\sum\limits_{i=n/2}^{n-1}}
\phi_{a_{2i}}\right)  ,
\end{align*}
which becomes
\[
p_{GB}=\sin^{n}\left(  ns\phi\right)
\]
under the condition that all azimuthal angles are of equal value $\phi
_{a_{2i}}=\phi$. The maximum violation bound is
\begin{equation}
p_{GB}^{\max}=1, \label{maxevenn}%
\end{equation}
when $\phi=\pi/\left(  2ns\right)  $.

The spin parity effect in the violation of GBI exists for the $n$-particle
entangled Schr\"{o}dinger-cat-state of spin-$s$ if the measurements are
restricted only in the subspace of SCSs. Moreover a particle number parity
effect is also demonstrated that the maximal violation bound is $p_{GB}^{\max
}=1/2$ for odd number and $p_{GB}^{\max}=1$ for even number.

\section{Conclusion}

We propose in this paper a GBI ($p_{GB}^{lc}\leq0$ Eq.(\ref{GBI})) for the
$n$-particle entangled Schr\"{o}dinger-cat-state of spin-$s$. It needs $n$
observers and total $2n-1$ measuring directions following the original BI with
$n=2$ and $s=1/2$. The GBI and its violation can be formulated in an unified
way by means of the SCS quantum probability statistics. The density operator
of entangled states is divided into local part and non-local part, which
describes the quantum interference of coherent superposition of entangled
multi-particle states. The local part leads to the GBI, while the non-local
part is responsible for the violation in quantum average.

For the $n$-particle entangled state of spin-$1/2$, the maximum violation
bound depends on the particle number that $p_{GB}^{\max}=1/2,1$ respectively
for odd and even $n$ in consistence with the known observation of larger
violation\cite{ex4} for even $n$. The GBI is never violated by the
$n$-particle entangled Schr\"{o}dinger-cat-state with higher spin-$s$ under
the quantum average. When the measuring outcomes are restricted to the
subspace of SCSs, namely only the maximum spin values $\pm s$ are taken into
account, the GBI is violated only by half-integer but not integer spin-$s$.
This spin parity effect\ is seen to be a direct result of Berry phase between
the SCSs of north- and south- pole gauges. The maximum violation bound of GBI
also depends on the particle number the same as the spin-$1/2$ case. The
particle number parity effect may have some applications in quantum
information associated with many-particle entanglement.

Our generic arguments of spin-parity effect could be tested specifically with
the photon-pairs generated by spontaneous parametric down conversion of light
through a BBO crystal. The two-particle entangled state of spin-$1/2$ is
easily demonstrated with the usual polarization entanglement of photons. And,
for the spin-$1$ entangled Schr\"{o}dinger-cat-state one needs to use the
orbital angular momentum entanglement\cite{prl2002}{, e.g., }$\left\vert
\psi\right\rangle _{1}=\left(  \left\vert l_{1}=1,l_{2}=1\right\rangle
+\left\vert l_{1}=-1,l_{2}=-1\right\rangle \right)  /\sqrt{2}$ of two Laguerre
Gaussian mode photons $LG_{\pm1}$ with the entanglement concentration
technique\cite{np2011}. The photon angular-momentum along arbitrary directions
$\pm\mathbf{a}_{1}$ and $\pm\mathbf{a}_{2}$ can be engineered and measured
independently by the detectors in each side respectively.

\section*{Acknowledgements}

This work was supported by the National Natural Science Foundation of China
(Grant Nos. 11275118, 11874247, 11904216 and 11974290 ).

\section*{Appendix}

\setcounter{equation}{0} \renewcommand\theequation{A\arabic{equation}}We
present the derivation of GBI for $n$-particle spin-$s$ entangled
Schr\"{o}dinger cat state in terms of hidden variable classical-statistics
following Bell\cite{Bell}. For the entangled Schr\"{o}dinger cat state of
$n=2$
\begin{equation}
|\psi\rangle=c_{1}|+s,+s\rangle+c_{2}|-s,-s\rangle, \label{A1}%
\end{equation}
with $\left\vert c_{1}\right\vert ^{2}+\left\vert c_{2}\right\vert ^{2}=1$,
the normalized measuring outcome values of two observers are denoted by
\begin{align*}
A_{1}(a_{1}\mathbf{)}  &  =\pm1,\\
A_{2}\left(  a_{2}\right)   &  =\pm1
\end{align*}
respectively along the measuring directions $a_{1}$ and $a_{2}$. The measuring
outcome correlation according to Bell\cite{Bell} is evaluated by the classical
statistics
\begin{align*}
p_{lc}\left(  a_{1},a_{2}\right)   &  =\int\rho\left(  \lambda\right)
A_{1}\left(  a_{1},\lambda\right)  A_{2}\left(  a_{2},\lambda\right)
d\lambda\\
&  \equiv\left\langle A_{1}\left(  a_{1}\right)  A_{2}\left(  a_{2}\right)
\right\rangle ,
\end{align*}
in which $\rho\left(  \lambda\right)  $ is the probability density
distribution\ of hidden variable\ $\lambda$. The product of two correlations
is%
\begin{align}
&  p_{lc}\left(  a_{1},a_{2}\right)  p_{lc}\left(  a_{2},a_{3}\right)
\nonumber\\
&  =\int\int\rho\left(  \lambda\right)  \rho\left(  \lambda^{\prime}\right)
A_{1}\left(  a_{1},\lambda\right)  A_{2}\left(  a_{2},\lambda\right)
A_{1}\left(  a_{2},\lambda^{\prime}\right)  A_{2}\left(  a_{3},\lambda
^{\prime}\right)  d\lambda d\lambda^{\prime}\nonumber\\
&  \leq\left\vert \int\int\rho\left(  \lambda\right)  \rho\left(
\lambda^{\prime}\right)  A_{1}\left(  a_{1},\lambda\right)  A_{2}\left(
a_{3},\lambda^{\prime}\right)  d\lambda d\lambda^{\prime}\right\vert
\nonumber\\
&  =\left\vert \left\langle A_{1}\left(  a_{1}\right)  \right\rangle
\left\langle A_{2}\left(  a_{3}\right)  \right\rangle \right\vert , \label{A2}%
\end{align}
since$\ A_{2}\left(  a_{2}\right)  =A_{1}\left(  a_{2}\right)  $ for the
parallel spin polarization of the entangled state Eq.(\ref{A1}) and $A_{2}%
^{2}\left(  a_{2}\right)  =1$. We define the classical probability
mean-deviation by%
\begin{align*}
\Delta A_{1}  &  \equiv A_{1}(a_{1})-\left\langle A_{1}(a_{1})\right\rangle
,\\
\Delta A_{2}  &  \equiv A_{2}(a_{3})-\left\langle A_{2}(a_{3})\right\rangle ,
\end{align*}
with $\left\langle A_{1}(a_{1})\right\rangle =\int\rho\left(  \lambda\right)
A_{1}(a_{1},\lambda)d\lambda$, and $\left\langle A_{2}(a_{3})\right\rangle
=\int\rho\left(  \lambda\right)  A_{2}(a_{3},\lambda)d\lambda$ being average
values of measuring outcomes. The average of deviation product is evaluated as%
\begin{align*}
\left\langle \Delta A_{1}\Delta A_{2}\right\rangle  &  =\left\langle
[A_{1}(a_{1})-\left\langle A_{1}(a_{1})\right\rangle ][A_{2}(a_{3})-\langle
A_{2}(a_{3})\rangle]\right\rangle \\
&  =\left\langle A_{1}(a_{1})A_{2}(a_{3})\right\rangle -\left\langle
A_{1}(a_{1})\right\rangle \left\langle A_{2}(a_{3})\right\rangle ,
\end{align*}
from which the right-hand side of Eq.(\ref{A2}) equals
\[
\left\vert \left\langle A_{1}(a_{1})\right\rangle \left\langle A_{2}%
(a_{3})\right\rangle \right\vert =\left\vert \left\langle A_{1}(a_{1}%
)A_{2}(a_{3})\right\rangle -\left\langle \Delta A_{1}\Delta A_{2}\right\rangle
\right\vert .
\]
Because $\left\langle A_{1}(a_{1})A_{2}(a_{3})\right\rangle $, $\left\langle
\Delta A_{1}\Delta A_{2}\right\rangle $ have the same sign and
\[
\left\vert \left\langle A_{1}(a_{1})A_{2}(a_{3})\right\rangle |\geq
|\left\langle \Delta A_{1}\Delta A_{2}\right\rangle \right\vert
\]
we have the inequality%
\begin{equation}
\left\vert \left\langle A_{1}(a_{1})\right\rangle \left\langle A_{2}%
(a_{3})\right\rangle \right\vert \leq\left\vert \left\langle A_{1}(a_{1}%
)A_{2}(a_{3})\right\rangle \right\vert . \label{A3}%
\end{equation}
Then Eq.(\ref{A2}) becomes%
\begin{align}
p_{lc}\left(  a_{1},a_{2}\right)  p_{lc}\left(  a_{2},a_{3}\right)   &
\leq\left\vert \left\langle A_{1}\left(  a_{1}\right)  \right\rangle
\left\langle A_{2}\left(  a_{3}\right)  \right\rangle \right\vert \nonumber\\
&  \leq\left\vert \left\langle A_{1}\left(  a_{1}\right)  A_{2}\left(
a_{3}\right)  \right\rangle \right\vert \nonumber\\
&  =\left\vert p_{lc}\left(  a_{1},a_{3}\right)  \right\vert . \label{A4}%
\end{align}
Also, the validity of the GBI Eq.(\ref{A4}) for $n=2$ can be easily verified
in terms of our quantum probability average with the local part of density
operator
\[
p_{lc}\left(  a_{1},a_{2}\right)  =\frac{1}{s^{2}}Tr\left[  \hat{\rho}%
_{lc}\hat{\Omega}(a_{1},a_{2})\right]  .
\]
The explicit forms of the local correlation probabilities for $s=1/2$ are
given by $p_{lc}\left(  a_{1},a_{2}\right)  =\cos\theta_{a_{1}}\cos
\theta_{a_{2}}$, $p_{lc}\left(  a_{2},a_{3}\right)  =\cos\theta_{a_{2}}%
\cos\theta_{a_{3}}$, and $p_{lc}\left(  a_{1},a_{3}\right)  =\cos\theta
_{a_{1}}\cos\theta_{a_{3}}$. We then have%
\begin{align}
&  p_{lc}\left(  a_{1},a_{2}\right)  p_{lc}\left(  a_{2},a_{3}\right)
\nonumber\\
&  =\cos\theta_{a_{1}}\cos^{2}\theta_{a_{2}}\cos\theta_{a_{3}}\nonumber\\
&  \leq|\cos\theta_{a_{1}}\cos\theta_{a_{3}}|\nonumber\\
&  =\left\vert p_{lc}\left(  a_{1},a_{3}\right)  \right\vert , \label{A5}%
\end{align}
which is valid in general for arbitrary three directions $a_{1},a_{2},a_{3}$
measured by two observers.

For $n=3$%
\[
\left\vert \psi\right\rangle =c_{1}|+s,+s,+s\rangle+c_{2}|-s,-s,-s\rangle,
\]
measuring outcome correlations for three observers along the directions
$\mathbf{a}_{1}$, $\mathbf{a}_{2}$, $\mathbf{a}_{3}$ is%
\begin{align*}
p_{lc}\left(  a_{1},a_{2},a_{3}\right)   &  =\int\rho\left(  \lambda\right)
A_{1}\left(  a_{1},\lambda\right)  A_{2}\left(  a_{2},\lambda\right)
A_{3}\left(  a_{3},\lambda\right)  d\lambda\\
&  \equiv\left\langle A_{1}\left(  a_{1}\right)  A_{2}\left(  a_{2}\right)
A_{3}\left(  a_{3}\right)  \right\rangle .
\end{align*}
The product of three correlations is%
\begin{align*}
&  p_{lc}\left(  a_{1},a_{2},a_{3}\right)  p_{lc}\left(  a_{2},a_{3}%
,a_{4}\right)  p_{lc}\left(  a_{3},a_{4},a_{5}\right) \\
&  =\int\int\int\rho\left(  \lambda\right)  \rho\left(  \lambda^{\prime
}\right)  \rho\left(  \lambda^{\prime\prime}\right)  \left[
\begin{array}
[c]{c}%
A_{1}\left(  a_{1},\lambda\right)  A_{2}\left(  a_{2},\lambda\right) \\
A_{3}\left(  a_{3},\lambda\right)  A_{1}\left(  a_{2},\lambda^{\prime}\right)
\\
A_{2}\left(  a_{3},\lambda^{\prime}\right)  A_{3}\left(  a_{4},\lambda
^{\prime}\right) \\
A_{1}\left(  a_{3},\lambda^{\prime\prime}\right)  A_{2}\left(  a_{4}%
,\lambda^{\prime\prime}\right) \\
A_{3}\left(  a_{5},\lambda^{\prime\prime}\right)
\end{array}
\right]  d\lambda d\lambda^{\prime}d\lambda^{\prime\prime}\\
&  \leq\left\vert
\begin{array}
[c]{c}%
\int\int\int\rho\left(  \lambda\right)  \rho\left(  \lambda^{\prime}\right)
\rho\left(  \lambda^{\prime\prime}\right)  A_{1}\left(  a_{1},\lambda\right)
A_{2}\left(  a_{3},\lambda^{\prime}\right) \\
\times A_{3}\left(  a_{5},\lambda^{\prime\prime}\right)  d\lambda
d\lambda^{\prime}d\lambda^{\prime\prime}%
\end{array}
\right\vert \\
&  =\left\vert \left\langle A_{1}\left(  a_{1}\right)  \right\rangle
\left\langle A_{2}\left(  a_{3}\right)  \right\rangle \left\langle
A_{3}\left(  a_{5}\right)  \right\rangle \right\vert ,
\end{align*}
since $A_{2}\left(  a_{2}\right)  =A_{1}\left(  a_{2}\right)  $, $A_{3}\left(
a_{3}\right)  =A_{1}\left(  a_{3}\right)  $, $A_{3}\left(  a_{4}\right)
=A_{2}\left(  a_{4}\right)  $ and $A_{i}^{2}\left(  a_{i}\right)  =1$. From
the inequality Eq.(\ref{A3}), it is easy to have%
\begin{align}
&  p_{lc}\left(  a_{1},a_{2},a_{3}\right)  p_{lc}\left(  a_{2},a_{3}%
,a_{4}\right)  p_{lc}\left(  a_{3},a_{4},a_{5}\right) \nonumber\\
&  \leq\left\vert \left\langle A_{1}\left(  a_{1}\right)  \right\rangle
\left\langle A_{2}\left(  a_{3}\right)  \right\rangle \left\langle
A_{3}\left(  a_{5}\right)  \right\rangle \right\vert \nonumber\\
&  \leq\left\vert \left\langle A_{1}\left(  a_{1}\right)  A_{2}\left(
a_{3}\right)  \right\rangle \left\langle A_{3}\left(  a_{5}\right)
\right\rangle \right\vert \nonumber\\
&  \leq\left\vert \left\langle A_{1}\left(  a_{1}\right)  A_{2}\left(
a_{3}\right)  A_{3}\left(  a_{5}\right)  \right\rangle \right\vert \nonumber\\
&  =\left\vert p_{lc}\left(  a_{1},a_{3},a_{5}\right)  \right\vert .
\label{A6}%
\end{align}

For arbitrary $n$%
\[
\left\vert \psi\right\rangle =c_{1}|+s\rangle^{\otimes n}+c_{2}|-s\rangle
^{\otimes n},
\]
we need total $2n-1$ independent measuring-directions labeled respectively by
$a_{1}$, $a_{2}$, $a_{3}$,..., $a_{2n-1}$ for $n$ observers. The product of
$n$ correlations leads directly the GBI that%
\begin{align*}
&  p_{lc}\left(  a_{1},a_{2},...,a_{n}\right)  p_{lc}\left(  a_{2}%
,a_{3},...,a_{n+1}\right)  p_{lc}\left(  a_{3},a_{4},...,a_{n+2}\right) \\
&  \times...p_{lc}\left(  a_{n},a_{n+1},...,a_{2n-1}\right) \\
&  =%
{\displaystyle\idotsint}
\left(
{\displaystyle\prod\limits_{i=1}^{n}}
\rho\left(  \lambda_{i}\right)  d\lambda_{i}\right)  \left[
\begin{array}
[c]{c}%
A_{1}\left(  a_{1},\lambda_{1}\right)  A_{2}\left(  a_{2},\lambda_{1}\right)
\\
A_{3}\left(  a_{3},\lambda_{1}\right)  ...A_{n}\left(  a_{n},\lambda
_{1}\right) \\
\times A_{1}\left(  a_{2},\lambda_{2}\right)  A_{2}\left(  a_{3},\lambda
_{2}\right) \\
A_{3}\left(  a_{4},\lambda_{2}\right)  ...A_{n}\left(  a_{n+1},\lambda
_{2}\right) \\
\times...\\
\times A_{1}\left(  a_{n},\lambda_{n}\right)  A_{2}\left(  a_{n+1},\lambda
_{n}\right) \\
A_{3}\left(  a_{n+2},\lambda_{n}\right)  ...A_{n}\left(  a_{2n-1},\lambda
_{n}\right)
\end{array}
\right] \\
&  \leq\left\vert
{\displaystyle\idotsint}
\left(
{\displaystyle\prod\limits_{i=1}^{n}}
\rho\left(  \lambda_{i}\right)  d\lambda_{i}\right)  \left[  A_{1}\left(
a_{1},\lambda_{1}\right)  A_{2}\left(  a_{3},\lambda_{2}\right)
...A_{n}\left(  a_{2n-1},\lambda_{n}\right)  \right]  \right\vert \\
&  =\left\vert \left\langle A_{1}\left(  a_{1}\right)  \right\rangle
\left\langle A_{2}\left(  a_{3}\right)  \right\rangle ...\left\langle
A_{n}\left(  a_{2n-1}\right)  \right\rangle \right\vert
\end{align*}
with $A_{i}\left(  a_{i}\right)  =A_{j}\left(  a_{i}\right)  ,$ $A_{i}%
^{2}\left(  a_{i}\right)  =1$. According to the\ Eq.(\ref{A3}), the above
equation becomes%
\begin{align}
&  p_{lc}\left(  a_{1},a_{2},...,a_{n}\right)  p_{lc}\left(  a_{2}%
,a_{3},...,a_{n+1}\right)  p_{lc}\left(  a_{3},a_{4},...,a_{n+2}\right)
\nonumber\\
&  \times...p_{lc}\left(  a_{n},a_{n+1},...,a_{2n-1}\right) \nonumber\\
&  \leq\left\vert \left\langle A_{1}\left(  a_{1}\right)  \right\rangle
\left\langle A_{2}\left(  a_{3}\right)  \right\rangle ...\left\langle
A_{n}\left(  a_{2n-1}\right)  \right\rangle \right\vert \nonumber\\
&  \leq\left\vert \left\langle A_{1}\left(  a_{1}\right)  A_{2}\left(
a_{3}\right)  ...A_{n}\left(  a_{2n-1}\right)  \right\rangle \right\vert
\nonumber\\
&  =\left\vert p_{lc}\left(  a_{1},a_{3},...,a_{2n-1}\right)  \right\vert .
\label{A7}%
\end{align}

\end{document}